\documentclass[aps,prb,frontmatterverbose,reprint,showpacs, superscriptaddress]{revtex4-1} 

\usepackage{graphicx}
\usepackage{dcolumn}
\usepackage{bm}
\usepackage{epstopdf}
\usepackage[colorlinks=true,linkcolor=blue, citecolor =blue]{hyperref}
\usepackage{multirow}

\usepackage[normalem]{ulem} 
\usepackage{color} 

\begin{document}
\newcommand{\uB}{$\mu_{B}$ }
\newcommand{\A}{$\mathrm{\mathring{A}}$}
\newcommand{\degree}{$^{\circ}$}
\newcommand{\AlO}{Al$_2$O$_{3}$}
\newcommand{\SiO}{SiO$_2$}
\newcommand{\AlOxx}{AlO$_x$}
\newcommand{\AlOx}{Al$_2$O$_{3-x}$}
\newcommand{\Swq}{$S(\omega,q)$}
\newcommand{\Sw}{$S(\omega$)}
\newcommand{\wbp}{$\omega_{BP}$ }	
\newcommand{\TiO}{TiO$_2$ }	
\title{Does the boson peak survive in an ultrathin oxide glass?}

\author{D.~L.\ Cortie}
\email {dcortie@uow.edu.au}
\affiliation{The Institute for Superconducting and Electronic Materials, University of Wollongong}
\author {M.~J. Cyster}
\affiliation{Chemical and Quantum Physics, School of Science, RMIT University, Melbourne, Victoria 3000, Australia}
\author{J.~S. Smith}
\affiliation{Chemical and Quantum Physics, School of Science, RMIT University, Melbourne, Victoria 3000, Australia}
\author{G. N. Iles}
\affiliation{Space Physics, School of Science, RMIT University, Melbourne, Victoria 3000, Australia}
\author{X. L. Wang}
\affiliation{The Institute for Superconducting and Electronic Materials, University of Wollongong}
\author{D. R. G. Mitchell}
\affiliation{Electron Microscopy Centre, University of Wollongong}
\author {R. A. Mole}
\author {N. de Souza}
\author {D. Yu}
\affiliation {Australian Nuclear Science and Technology Organisation}
\author{J.~H. Cole}
\email {jared.cole@rmit.edu.au}
\affiliation{Chemical and Quantum Physics, School of Science, RMIT University, Melbourne, Victoria 3000, Australia}

	\date{29 July, 2019}
	
\begin{abstract}
Bulk glasses exhibit extra vibrational modes at low energies, known as the boson peak. The microscopic dynamics in nanoscale alumina impact the performance of qubits and other superconducting devices, however the existence of the boson peak in these glasses has not been previously measured. Here we report neutron spectroscopy on Al/\AlOx\ nanoparticles consisting of spherical metallic cores from 20 to 1000 nm surrounded by a 3.5 nm thick alumina glass. An intense low-energy peak is observed at $\omega_{BP}$ = 2.8 $\pm$ 0.6 meV for highly oxidised particles, concurrent with an excess in the density of states. The intensity of the peak scales inversely with particle size and oxide fraction indicating a surface origin, and is red-shifted by 3 meV with respect to the van-Hove singularity of $\gamma$-phase \AlO\ nanocrystals. Molecular dynamics simulations of $\alpha$-\AlO, $\gamma$-\AlO\ and a-\AlOx\ show that the observed boson peak is a signature of the ultrathin glass surface, and the frequency is softened compared to that of the hypothetical bulk glass. 
\end{abstract} 

\maketitle

Amorphous materials make up a large fraction of terrestrial matter and their associated glass transitions have consequences in the science of minerals and electronic materials.\cite{Novikov2004} Despite the structural differences between oxide, metal and polymer glasses, there are striking universal features in the dynamic properties of disordered solids.\cite{Loidl_l989} These include an enhancement in the vibrational density of states at terahertz frequencies (THz) and a corresponding plateau in the heat conductivity.\cite{Loidl_l989} The latter results in an enigmatic feature in Raman and neutron spectroscopy observable in the THz (meV) frequency range called the ``boson peak''. One characteristic length scale is defined by the ratio of the speed of sound ($v$) to the boson peak frequency $v_{wp}$ and falls in the range $l_c \approx$ 0.3--5~nm for most glasses.\cite{Hassan1994} In oxides, it has become clear that the glass boson peak is correlated with the acoustic phonon van-Hove singularity (vHs) in related crystalline networks, but is red-shifted by several meV,\cite{Dove1997,Hiroshi2008,Nakayama_2002,Binder2011,Chumakov2011} probably reflecting the level of elastic disorder over the characteristic spatial scale $l_c$.\cite{Ganter1998} The nature of the boson peak in nanostructures is not well known, however interfaces may create a strong source of inhomogeneity at a spatial scale comparable to $l_c$ and could shift or broaden the peak considerably. While a well-defined boson peak has been observed at the surface of a semi-infinite solid,\cite{Steurer2007,Steurer_2008} the situation is less clear for ultrathin nanostructures. 

Nanoglasses are far less studied than their bulk analogues. They are, however, widely used as junction layers in electronics and quantum computing, and are also ubiquitous in nature because thermodynamics often favours forming stable (or metastable) glasses in nanostructures. Whereas bulk glass-formers are rare, the majority of metal oxides can exhibit amorphous phases as thin film layers. For example, alumina \AlOx\ glass is generally only stable as a thin surface layer, unlike the canonical glass \SiO\ which is an intrinsic bulk glass-former\cite{Lee2009}. The lack of a bulk alumina glass means that, to the best of our knowledge, the frequency and very existence of the boson peak has not been measured until now.

This basic question has taken on an applied dimension with the development of superconducting electronics based on thin Al/\AlOxx/Al junctions. The self-limiting nature of the oxidation of aluminium provides an ideal fabrication method for producing uniform, nanoscale insulating layers which are the key to SQUIDs\cite{Fagaly2006} and superconducting qubits.\cite{Clark2008,Devoret1169, devoret2004,Wendin2007} Yet a significant drawback of this material is that two-level systems dominate the low temperature physics of glasses below 1 K,\cite{Phillips_1987} 
and limit the coherence, fidelity and reproducibility of superconducting devices.\cite{Muller2017, Paladino2014} It is therefore essential to understand the microscopic degrees of freedom in nanoscale glasses in order to pinpoint the mechanisms that produce this noise.

\begin{figure*}[t!]
  \includegraphics[width=1.95\columnwidth]{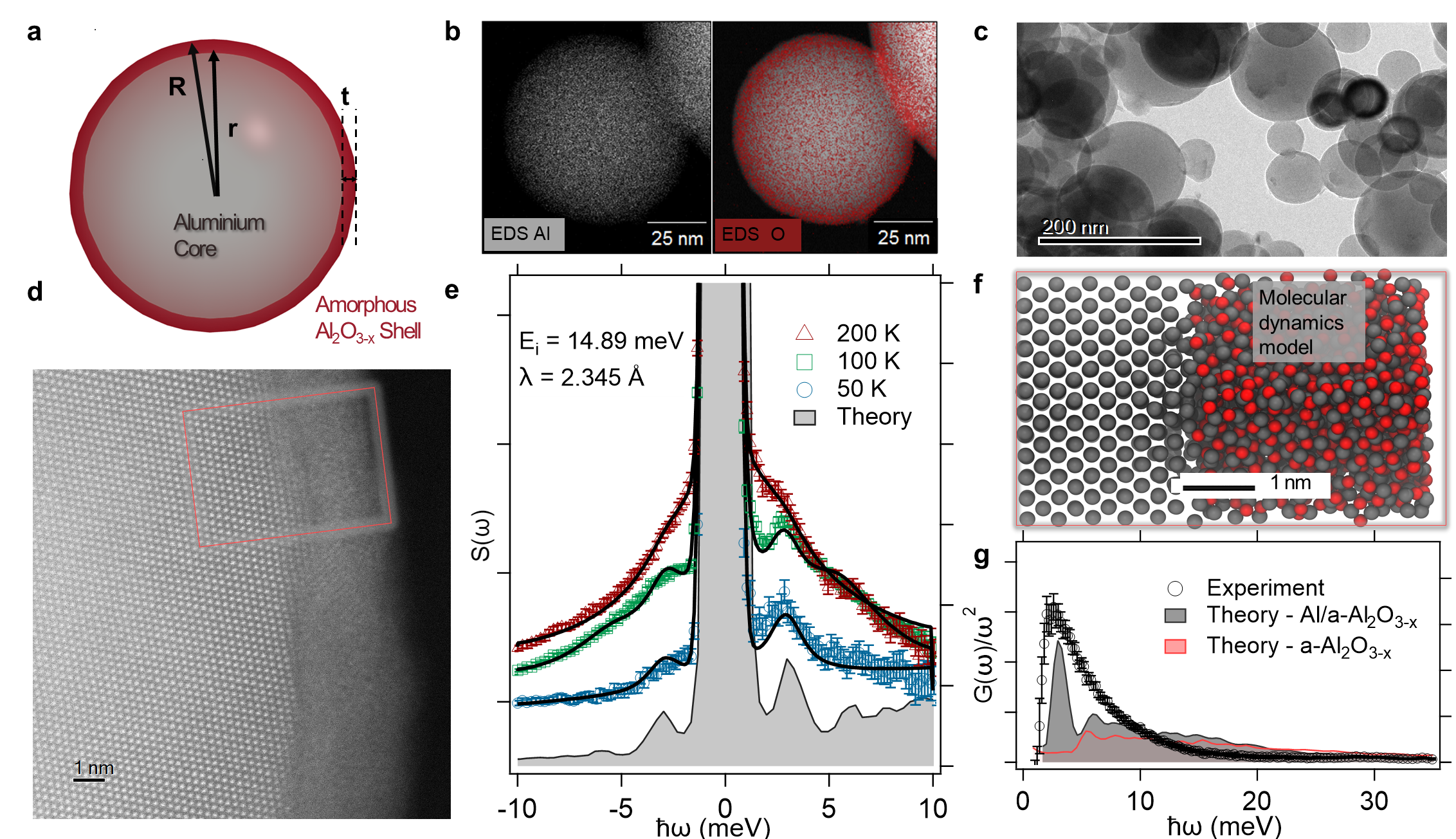}
  \caption{a) Schematic illustration of the core-shell structure formed by the native oxide on aluminium nanoparticles described by an outer radius $R$, an inner radius $r$ and a thickness $t$. b) Energy-dispersive X-ray spectroscopy (EDS) maps of the aluminium profile in spherical particles and oxygen profile surrounding the particles. c) Low magnification TEM image of the ensemble of spherical Al/\AlOx\ particles. d) High angle angular darkfield (HAADF) scanning TEM image of the interface between the crystalline core and oxide shell e)  Inelastic neutron scattering intensity $S(\omega)$ at several temperatures. The solid lines are fits including the detailed balance factor. The theoretical spectra calculated from the MD calculations is offset below for clarity. f) Snapshot of a MD configuration of the 3~nm thick glass interface on a crystalline substrate. g)  Generalized density of states measured by inelastic neutron spectroscopy for $\left<R\right>$ = 35 nm particles at 100 K scaled by $w^{-2}$. The shaded regions are those calculated from the MD of a Al/\AlOx\ interface and a hypothetical bulk \AlOx.}
  \label{fig:FigOne}
\end{figure*}

To measure the picosecond dynamics in ultrathin alumina, we performed inelastic neutron spectroscopy on well-characterized Al/\AlOx\ structures (Fig.~\ref{fig:FigOne}). Neutron spectroscopy is the ideal tool to study atomistic dynamics in complex structures because it offers both energy and (reciprocal) spatial resolution. Time-of-flight spectroscopy is particularly well suited for studying the spectra of polycrystalline nanopowders \cite{Cortie_2019,Saviot2008,Deniz2016} and glasses.\cite{Nakayama_2002} Here we deployed the PELICAN instrument \cite{Yu2013} -- a cold neutron time-of-flight spectrometer at the Australian Nuclear Science and Technology Organisation -- operating at wavelengths of 4.69 \A\ and 2.345 \A. Spherical aluminium nanoparticles were supplied by US Research Nanomaterials, where they were produced by a high temperature electrical explosive method. The particles were deliberately exposed to ambient atmosphere to spontaneously form ultrathin amorphous oxide skins which are well-known to passivate aluminium surfaces. This results in a core-shell nanoparticle (Fig.~\ref{fig:FigOne}a) where $t$ is the thickness of the oxide skin, and $R$ is the total radius. Past studies reported that the surface oxide has a thickness between 2--4 nm which varies slightly for different aluminium facets. \cite{Evertsson2015, Reichel2008} This was directly confirmed in the spherical particles using aberration-corrected transmission electron microscopy for the exposed Al samples (Fig.~\ref{fig:FigOne}b), indicating the oxide shell has $t = $ 3.5 $\pm$ 0.5 nm with excellent uniformity across the different particles. Three nanopowder samples were measured, each consisting of an ensemble of spheres (e.g. Fig.~\ref{fig:FigOne} c) characterised by different average core sizes, $\left<R\right>$ = 35 $\pm$ 15, 45 $\pm$ 21 and 350 $\pm$ 100 nm, where the error bar is the measured standard deviation. Atomic resolution imaging shows that the aluminium core is crystalline, whereas the external region is amorphous, as can be seen in the high angular-annular-dark-field (HAADF) image in Fig.~\ref{fig:FigOne}d. The volume fraction of the glass region surrounding the metal core is given by: $V_F = (t^3 + 3R^2t - 3Rt^2)/ {R^3} $. In smaller particles where $R \approx$ 20--60 nm, a substantial fraction of the overall mass is in the oxide skin (20--70\%). The coherent neutron scattering cross-section (scattering length) for oxygen ($b_O$ = 5.803 fm) is nearly twice that of aluminium ($b_{Al}$ = 3.45 fm). Thus, the oxide signal is further enhanced by a factor of $ \frac{\rho_{AlO}(b_O + b_{Al})}{\rho_{AlO}(b_O)}\frac{M_{Al}}{M_{AlO}} \approx 1.9$ where $\rho$ is the mass density and $M$ is the molar mass. Consequently, in fine powders, we resolved an intense feature in the q-integrated neutron scattering intensity \Sw\ (as shown in Fig.~\ref{fig:FigOne}e) which is dominated by the oxide fraction. The feature is very similar to the spectral signature of the boson peak in bulk amorphous \SiO,\cite{Nakayama_2002} however the frequency is softened by 40\% which may indicate that the alumina glass is more fragile.\cite{Hassan1994} By using neutrons with an incident energy of 14.89 meV, the feature can be resolved in both the neutron energy-loss and energy-gain sides of the elastic signal, thereby allowing for observation of the temperature-dependent asymmetry. As shown by the solid black lines in Fig.~\ref{fig:FigOne}e, the spectra can be fitted using a combination of two Lorentzians and a Gaussian, weighted by the thermal balance factor. The latter indicates the spectra obeys Bose-Einstein statistics, a key signature of a bosonic excitation common to all glass boson peaks. 

\begin{figure}
  \includegraphics[width=0.9\columnwidth]{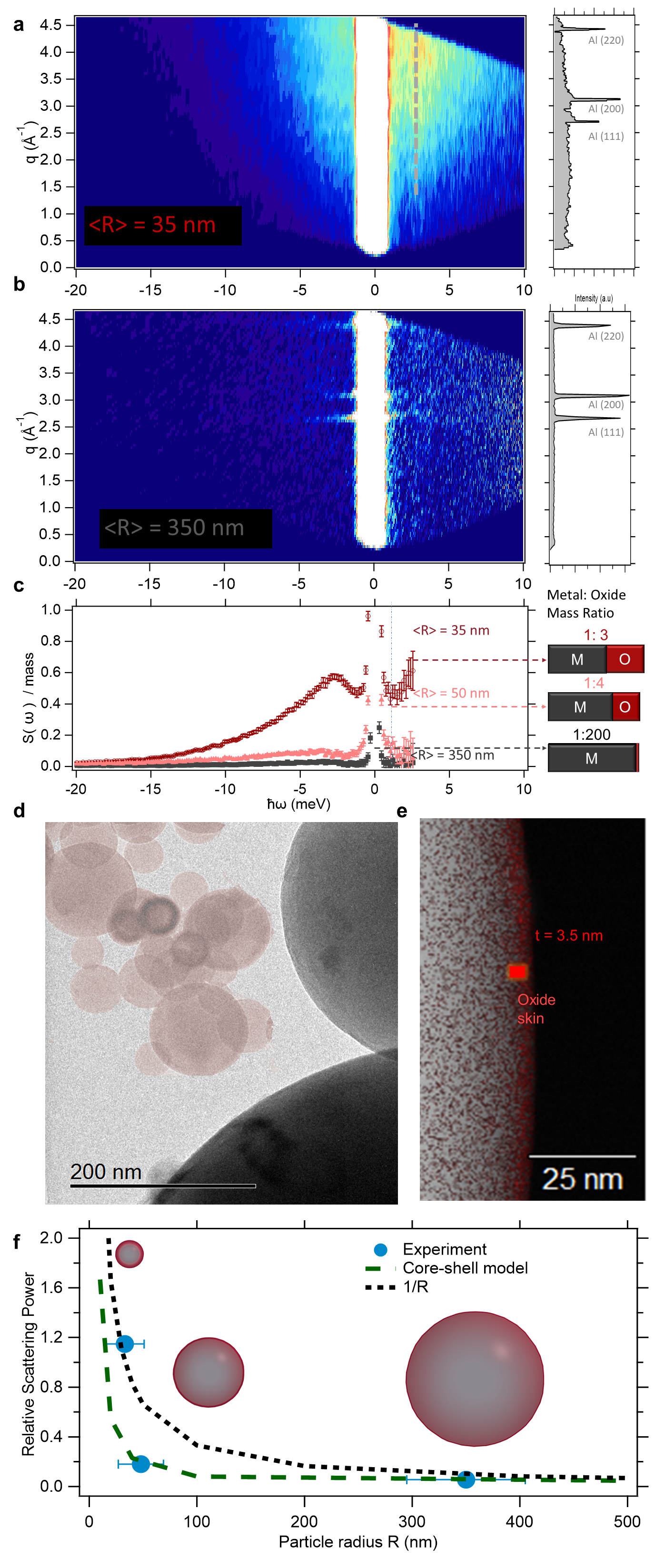}
  \caption{a) \Swq\ map for the $\left<R\right>$ = 35 nm Al nanopowder showing the $q$-dependency of the excitation with the diffraction pattern (right) extracted from the elastic channel. b) \Swq\ map for the $\left<R\right>$ = 350 nm with the  ($E$ = 0) diffraction pattern (right). c) Q-integrated S($\omega$) normalized by sample mass for the different Al/\AlOx\ nanopowders.
 d) TEM image for the $\left<R\right>$ = 35 nm Al nanoparticles (red) superimposed on an image of the $\left<R\right>$ = 350 nm particles (grey). e) EDS map showing the oxide thickness of 3.5 nm on a large particle from the $\left<R\right>$ = 350 nm ensemble (left). f) The intensity of \wbp\ feature scaled as a function of average particle radius, compared with the theoretical scattering power of the shell.}
  \label{fig:FigTwo}
\end{figure}

Molecular dynamics (MD) simulations predict a similar low energy feature in alumina glass. We calculated atomic trajectories for bulk $\alpha$-, $\gamma$-, and amorphous-Al$_2$O$_3$, and Al/AlO$_x$ surfaces. Each simulation was run for a total of 50 ps where the equations of motion were solved every 1 fs using the General Utility Lattice Program (GULP).\cite{Gale1997} An example of the crystal-glass Al/\AlOx\ surface model is shown in Fig.~\ref{fig:FigOne}f. The force-fields between atoms were modelled via the Streitz-Mintmire empirical potential \cite{Streitz1994} which includes an electronegativity correction. \cite{Mortier1986} The intermediate neutron scattering function was calculated from the spatial coordinates of the MD trajectories, and Fourier transformed to get the simulated scattering function \Sw\ using NMoldyn.\cite{Hinsen2012} Figure \ref{fig:FigOne}e shows the calculated spectra for a 3 nm oxide skin interface on an aluminium substrate at 100 K, where the main features are well captured by the model. The vibrational density of states $G(\omega)$ calculated from the velocity autocorrelation functions, normalized to Debye units scaled by $\omega^{-2}$ is shown in Fig.~\ref{fig:FigOne}g. This exhibits an excess of modes centred on \wbp\ = 2.8 $\pm$ 0.6 meV, as found for boson peaks in glasses, and vHs in crystals (Fig.~\ref{fig:FigOne}e). While both pure amorphous \AlOx\ and Al/\AlOx\ MD simulations predict an excess density of states between 0--30 meV, the interface models predict that the first features in the ultrathin oxide have a lower frequency than in the hypothetical bulk glass, matching well with experiment.

According to the $q$-dependent neutron scattering measurements, the feature is dispersionless unlike typical acoustic phonon modes, but similar to all glass boson peaks. Figure \ref{fig:FigTwo}a shows the full \Swq\ powder-averaged neutron scattering map for small ($\left<R\right>$ = 35 nm) particles in terms of the reciprocal space vector magnitude ($q$) and the energy transfer. In the right section, the diffraction pattern extracted from the zero energy (elastic) slice is plotted. Although the nanoglass is easily concealed in the diffraction measurements, which are dominated by the crystalline aluminium peaks, it shows up as a strong feature in the inelastic neutron spectroscopy with a peak at $\omega_{BP}$ = 2.8 $\pm$ 0.6 meV. The central frequency of the low energy feature is  independent of $q$, indicating a quasi-localized mode. The $q$-width of the peak is broad, and not sharply modulated at the metal's Bragg points, indicating it originates from a different spatial scale. Indeed, the scattering intensity is peaked towards 4.0 $\mathrm{\mathring{A}}^{-1}$ which is the characteristic Al-O distance in the glass tetrahedra. In common with the boson peak in \SiO, the inelastic signal peaks at the second elastic maxima in the glass static structure factor because the first maxima in $S(q)$ corresponds to length-scales of connected tetrahedra that are less rigid than one tetrahedra leading to a less pronounced peak.\cite{Binder2011}. For the largest particles ($\left<R\right>$ = 350 nm) the feature is only barely detectable when plotted on the same scale and normalised by mass (Fig.~\ref{fig:FigTwo}b), however the aluminium diffraction peaks are sharper and stronger. This is consistent with the tiny oxide fraction ($<1\%$) in the bigger particles which is expected as the oxide skin has the same uniform thickness, whereas the aluminium cores are very different sizes (Fig.~\ref{fig:FigTwo}c,d). Theoretically, the thickness of the skin is determined by the thermodynamic stability criteria of the oxide which depends on surface energy and tends to disfavour growth beyond 3--4 nm.\cite{Tavakoli2013,Jeurgen2000} Figure \ref{fig:FigTwo}e shows an EDS map of the oxide skin on a larger particle indicating that, as in the small particles, $t$ = 3.5 nm. The different shell:core ratios explains why the intensity of observed feature scales as summarized in Fig.~\ref{fig:FigTwo}f. The scaling for a core-shell particle is subtly different from a straightforward 1/$R$ scaling. We also noted small $\pm 1 $ meV shifts in $\omega_{BP}$ between the various samples indicating that strain caused by surface curvature may slightly influence the boson peak frequency since small-sized nanoparticles are under an effective pressure due to surface tension.\cite{Chatterji2010}

\begin{figure}
  \includegraphics[width=0.93\columnwidth]{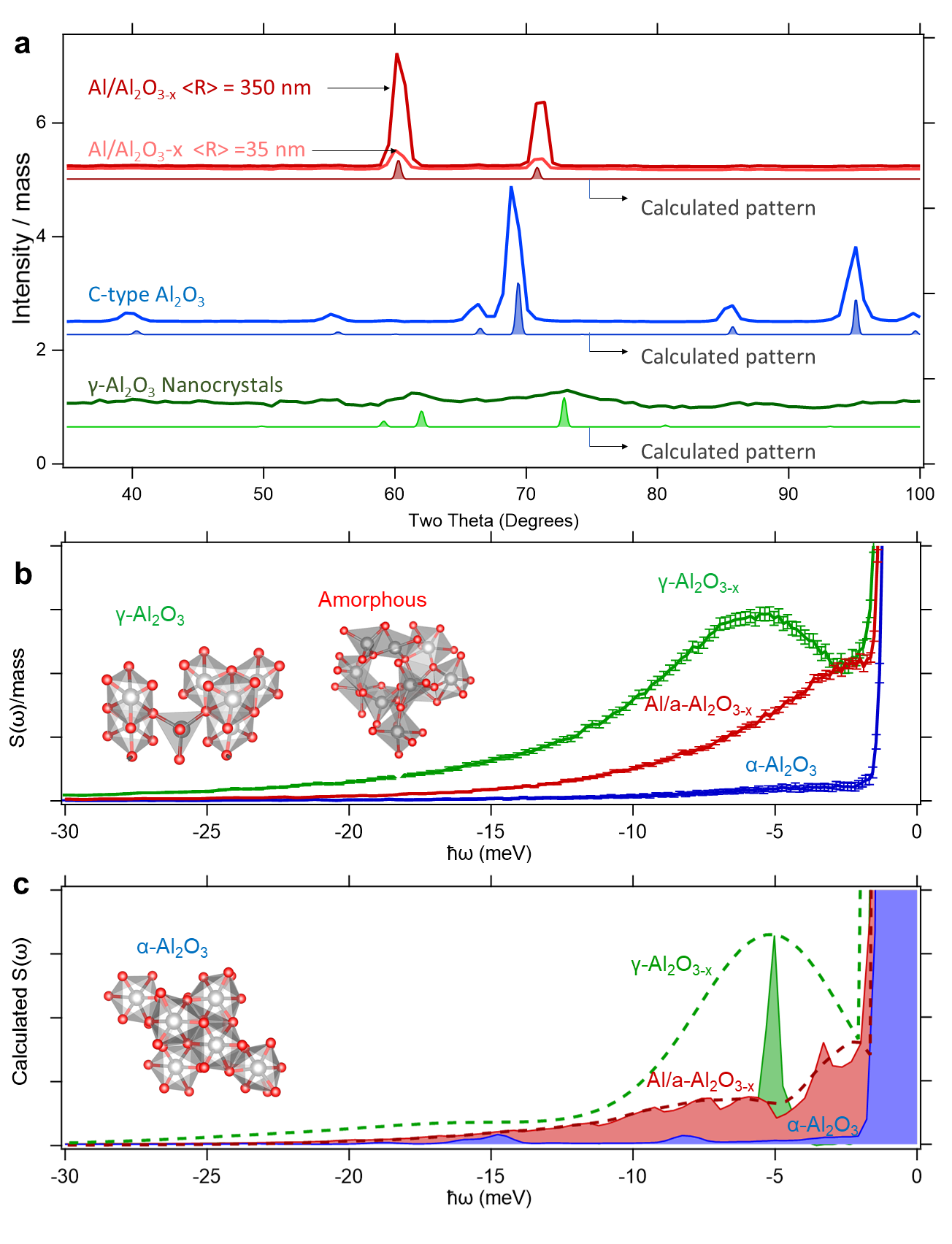}
  \caption{a) Neutron diffraction patterns at $\lambda = 2.35$ \A\ for the Al/\AlOx, $\gamma$- and $\alpha$-\AlO\ taken from the elastic channel.  
    b) q-integrated inelastic neutron scattering intensity for $\gamma$-\AlO\ and $\alpha$-\AlO\ nanopowders compared with the Al/\AlOx\ coreshell particles, normalized by sample mass. c) Simulated \Sw\ using classical force-field MD simulations for infinite periodic cells of $\gamma$-\AlO, $\alpha$-\AlO\, and the a-Al/\AlOx\ interface. The shaded regions assume the instrument resolution of 1 meV, whereas the dashed lines indicate the effect of including broadening in the simulations from finite-size or disorder. The insets are snapshots from  small sections of the  MD models showing the different tetrahedral and octahedral networks. }
  \label{fig:FigThree}
\end{figure}

Although the data for nanoalumina display striking similarities to bulk glasses, it would be remiss if we failed to note that other spectroscopic studies of nanopowders have reported anomalous boson-peak features. These have been observed even in materials such as nanocrystalline Fe \cite{Fultz1997} and TiO$_2$ \cite{Saviot2008} where no such feature would be expected. Those studies did not attribute this to an interfacial glass. One alternate proposal was that weak coupling between nanocrystallites in a non-dispersed powder leads to a ``microstructural'' boson peak analogous to disordered solids. \cite{Fultz2004} To explore this, we performed control measurements on other nanocrystalline aluminium oxides all of which had a similar, low packing density ($25\%$).  Figure \ref{fig:FigThree}a displays the diffraction patterns for the different nanosized aluminium and aluminium oxide samples, along with the calculated patterns from the Inorganic Crystal Structure Database (ICSD). The corundum-type and aluminium cores are of good crystal quality, evident in the strong Bragg peaks, whereas the $\gamma$ particles are only barely crystalline (see Supplemental Information). Figure \ref{fig:FigThree}b compares the inelastic neutron spectroscopy for various samples all measured under identical conditions. It is clear that the spectra are very different despite having similar packing densities. The corundum phase shows almost no features below 10 meV, whereas both the $\gamma$ and Al/\AlOx\ particles have strong low energy features. This agrees with past calculations and measurements in that the first vHs in corundum is at much higher energy (30 meV),\cite{Heid2000} whereas pure $\gamma$ oxide has its first vHs at 7 meV as predicted by lattice dynamics calculations.\cite{Ching_2008} The differences reflect the characteristics of the different polyhedra networks. The insets of Fig.~\ref{fig:FigThree} are snapshots of the tetrahedra and octahedra in the MD of the various phases. The $\gamma$-phase has more in common with the amorphous phase in that it has similar bond lengths and features a large fraction of tetahedral units (40\%).\cite{Ching_2008} Unlike the octahedral units in corundum, corner-sharing tetrahedra are not strongly constrained thus offering low energy modes of distortion. The alumina glass has an average Al-O coordination number of 4.5 with a larger fraction of corner-sharing tetrahedra. Therefore, the boson peak in glassy alumina, which is red-shifted with respect to the vHs of crystalline $\gamma$ phase, has close analogies with the boson peak in amorphous silica, which is red-shifted with respect to the vHS of crystalline crystobalite.\cite{Dove1997} The observed features in the crystalline solids also match with the MD calculations. Figure \ref{fig:FigThree}c shows the theoretical neutron spectra calculated from the MD for periodic boundary conditions, ignoring inter-nanoparticle interactions. While the primary frequencies match, the experiment shows considerable broadening from disorder or finite-size effects. These are real features because the broadening exceeds 1 meV, whereas the instrument resolution is 0.14 - 0.6 meV .\cite{Yu2013} It has been proposed that interfacial-scattering reduces phonon (and magnon) lifetimes in nanocrystals leading to line broadening of the order $\Gamma = h \frac{2v}{L}$ where $v$ is the speed of sound and $L$ is the size of the particle/domain.\cite{Cortie_2019,Bayrakci_2013} Assuming the speed of sound to be 9 km/s, $\Gamma$ is 1 meV for a crystallite with $L=70$ nm. Experimentally, the broadening is closer to 1.5 meV in the aluminium particles and 6 meV in the $\gamma$-\AlO\ consistent with the smaller particle sizes in the latter (Supplemental). The dashed lines in Fig.~\ref{fig:FigThree}c show the effect of including line-width broadening for the simulated spectra. As the boson peak for the Al/\AlOx\ particles has a low dispersion, $v \approx 0$, it does not appear to be broadened in the same way as the ordinary phonon features. 

In conclusion, all available experimental and theoretical evidence points to the existence of distinctive boson peak features in quasi-2D ultrathin alumina glasses which contribute excess low energy vibrational modes. Measuring the dynamics of the Al/\AlOx\ is an important first step in understanding the microscopic degrees of freedom that lead to the two-level system noise in qubits at low temperature.

\textbf{Acknowledgements:}
JC and DC acknowledge the support of the Australian Research Council(ARC) via DP140100375, DE180100314. This work was partly supported by the ARC Centre for Excellence in Future Low-Energy Electronics (CE170100039) and the Centre for Excellence in Exciton Science (CE170100026). High performance computing performed on the National Computational Infrastructure (NCI). This research used the JEOL JEM-ARM200F funded by the ARC LIEF grant (LE120100104). Neutron spectroscopy was performed at the Australian Centre for Neutron Scattering (P7437).

\bibliographystyle{apsrev4-1}

\end{document}